# Enhancing Automatic PT Tagging for MEDLINE Citations Using Transformer-Based Models

Victor H. Cid*, James Mork

National Library of Medicine, Bethesda, Maryland, US

* Corresponding author: cidv@mail.nlm.nih.gov

**Abstract**

We investigated the feasibility of predicting Medical Subject Headings (MeSH) Publication Types (PTs) from MEDLINE citation metadata using pre-trained Transformer-based models BERT and DistilBERT. This study addresses limitations in the current automated indexing process, which relies on legacy NLP algorithms. We evaluated monolithic multi-label classifiers and binary classifier ensembles to enhance the retrieval of biomedical literature. Results demonstrate the potential of Transformer models to significantly improve PT tagging accuracy, paving the way for scalable, efficient biomedical indexing.

**Keywords:** MEDLINE, MeSH Publication Types, Pre-trained Foundation Models, Natural Language Processing, Machine Learning

## 1. Introduction

The MEDLINE indexed subset of the National Library of Medicine's (NLM's) PubMed service is a cornerstone of biomedical knowledge, housing millions of citations from journals worldwide. Its significance lies not only in its vast scope but also in its ability to organize and provide efficient access to this wealth of information. MEDLINE's indexing accuracy directly impacts knowledge dissemination. By enabling users to search biomedical citations using MeSH Publication Types (PTs) [1], MEDLINE empowers researchers, clinicians, and policymakers to navigate this resource with precision and purpose. Accurate PT tagging is essential for maintaining this functionality, and while current automated processes using legacy NLP algorithms effectively provide the desired baseline performance, there are opportunities to enhance their precision and scalability to meet evolving demands. This research explores whether exploiting transfer-learning techniques with Transformer-based machine learning models can enhance the efficiency and accuracy of PT prediction from MEDLINE metadata, addressing performance challenges and advancing the field of automated indexing.

## 2. Previous work

### 2.1. MEDLINE and PT Tags

MEDLINE organizes biomedical citations with MeSH PTs [1] [2], facilitating targeted searches. PTs categorize documents by research methodologies and publication formats, such as "Clinical Trials" or "Review," rather than content. Previous work has shown the utility of MEDLINE Publication Type tags to perform a variety of literature screening tasks (e.g., [3], [4], [5]).

## 2.2. Existing Automated Systems

The NLM Medical Text Indexing (MTI) project was a pioneering effort to enhance the efficiency and accuracy of indexing biomedical literature within MEDLINE [6] [7], culminating in a fully automated system that uses natural language processing (NLP) and machine learning algorithms to assign Medical Subject Headings (MeSH) terms to articles [8] [9] [10]. The NLM Medical Text Indexing project has utilized legacy NLP models to automate PT tagging. While effective, these systems lack robustness against the growing complexity of biomedical publications [11]. Table 1 shows examples of the MTI system's performance for a subset of PTs supported by the system. Macro-averages of precision, recall and f1-score across the PTs supported by the MTI application in their sample were 0.84, 0.53 and 0.64, respectively.

| Publication Type | Precision | Recall | F-1 score |
| --- | --- | --- | --- |
| Address | 1.00 | 0.25 | 0.40 |
| Case Reports | 0.90 | 0.49 | 0.64 |
| Clinical Trial | 0.25 | 0.01 | 0.02 |
| Clinical Trial Protocol | 0.49 | 0.48 | 0.48 |
| Clinical Trial, Phase I | 0.75 | 0.39 | 0.51 |
| Clinical Trial, Phase II | 0.81 | 0.48 | 0.60 |
| Clinical Trial, Phase III | 0.84 | 0.49 | 0.62 |
| Clinical Trial, Phase IV | 1.00 | 0.18 | 0.31 |
| Congress | 1.00 | 0.09 | 0.16 |
| Controlled Clinical Trial | 0.11 | 0.05 | 0.07 |
| Equivalence Trial | 0.50 | 0.07 | 0.13 |
| Historical Article | 0.99 | 0.48 | 0.65 |
| Interview | 1.00 | 0.17 | 0.29 |
| Meta-Analysis | 0.83 | 0.86 | 0.85 |
| Multicenter Study | 0.81 | 0.33 | 0.47 |
| Observational Study | 0.82 | 0.45 | 0.58 |
| Observational Study, Veterinary | 0.33 | 0.67 | 0.44 |
| Practice Guideline | 0.50 | 0.03 | 0.05 |
| Pragmatic Clinical Trial | 0.50 | 0.10 | 0.17 |
| Randomized Controlled Trial | 0.81 | 0.71 | 0.76 |
| Randomized Controlled Trial, Veterinary | 0.36 | 0.45 | 0.40 |
| Review | 0.80 | 0.55 | 0.65 |
| Systematic Review | 0.93 | 0.80 | 0.86 |
| Twin Study | 0.50 | 0.12 | 0.19 |
| Video-Audio Media | 0.46 | 0.08 | 0.14 |

*Table 1: PT Prediction performance of the automated MTI PT system for selected PTs from among those supported by the system, calculated from a sample of 40K MEDLINE citations. Data kindly supplied by the NLM MTI team.*

Concluding in 2024, the MTI project significantly expedited the indexing process, with subsequent automated efforts delegated to other departments within a restructured NLM organization.

## 2.3. Advancements in NLP

Transformer-based models, such as BERT, DistilBERT, and domain-specific adaptations like BioBERT, have revolutionized text classification tasks with their contextual understanding and transfer learning capabilities. BERT and DistilBERT are pre-trained on an extensive corpus of 3.3 billion words, including science topics, and are equipped to understand a large vocabulary and capture a wide range of language patterns and nuances. Several studies have successfully tested the feasibility of using foundation models to infer specific Publication Types from published literature. The studies have focused on individual publication types of particular interest, such as "Randomized Controlled Trials" [12] [13] and "Systematic Reviews" [14] [15]. The efficacy of transformer-based models for document classification has been demonstrated in various NLP tasks [16] [17] [18] [19]. Studies have also demonstrated the value of carefully preprocessing text data in preparation for machine learning-based text classification tasks [20] [21] [22].

Multiple studies have demonstrated the utility of BERT and reduced models such as DistilBERT and model ensemble approaches to accomplish scientific text classification and other related NLP tasks [23] [24] [25] [26] [27] [28]. Reported advantages of foundation models include improved performance, reduced need for feature engineering, easier training via transfer learning, effective contextual understanding, and the potential for continuous improvement, among others.

## 3. Challenges in PT Prediction

Accurate prediction of MeSH Publication Types (PTs) from MEDLINE citation metadata presents several significant challenges, which can be broadly categorized into three groups:

Data-Related Issues: A significant challenge lies in the characteristics of the data itself. Our corpus consists of over 5 million MELDINE citations manually indexed between 2016 and 2023; the indexing process uses a large number of PT labels (see Table 2), which complicates label prediction. Extreme class imbalance is prevalent, with common tags like "Journal Article" dominating the dataset, while rare tags such as "Adaptive Clinical Trial" are scarcely represented. This imbalance leads to biased models that perform well on frequent classes but struggle with rare ones. Table 2 shows the number of citations with each PT tag in our data.

| Publication Type | Count | Publication Type | Count | Publication Type | Count |
|---|---|---|---|---|---|
| Journal Article | 5050200 | Biography | 11967 | Webcast | 736 |
| Research Support, Non-U.S. Gov't | 1664856 | Clinical Trial, Phase III | 9777 | Clinical Conference | 733 |
| Review | 636047 | Practice Guideline | 8125 | Lecture | 704 |
| Research Support, N.I.H., Extramural | 427682 | Clinical Trial, Phase I | 8018 | Classical Article | 699 |
| Case Reports | 315324 | Portrait | 7108 | Address | 688 |
| Comparative Study | 196357 | Clinical Trial Protocol | 5834 | Legal Case | 464 |
| Randomized Controlled Trial | 144912 | Comment | 5757 | Research Support, American Recovery and Reinvestment Act | 326 |
| Research Support, U.S. Gov't, Non-P.H.S. | 140253 | Controlled Clinical Trial | 4872 | Observational Study, Veterinary | 264 |
| Letter | 123266 | Clinical Study | 4444 | Festschrift | 170 |
| Multicenter Study | 121202 | Personal Narrative | 4290 | Published Erratum | 164 |
| Systematic Review | 107409 | Congress | 3985 | Bibliography | 113 |
| Observational Study | 98699 | Interview | 3566 | Newspaper Article | 70 |
| Editorial | 87313 | Randomized Controlled Trial, Veterinary | 3073 | Consensus Development Conference, NIH | 67 |
| Meta-Analysis | 83454 | Retracted Publication | 2954 | Interactive Tutorial | 48 |

| Evaluation Study | 51307 | Consensus Development Conference | 2711 | Adaptive Clinical Trial | 35 |
| --- | --- | --- | --- | --- | --- |
| Clinical Trial | 39541 | Twin Study | 2422 | Corrected and Republished Article | 29 |
| Historical Article | 38560 | Pragmatic Clinical Trial | 1866 | Scientific Integrity Review | 19 |
| Validation Study | 34530 | Clinical Trial, Veterinary | 1827 | Directory | 17 |
| Video-Audio Media | 24506 | Dataset | 1414 | Duplicate Publication | 15 |
| Research Support, U.S. Gov't, P.H.S. | 22643 | Clinical Trial, Phase IV | 1205 | Periodical Index | 11 |
| Introductory Journal Article | 20279 | Patient Education Handout | 1094 | Retraction of Publication | 10 |
| Research Support, N.I.H., Intramural | 20250 | Equivalence Trial | 998 | Dictionary | 7 |
| News | 17514 | Guideline | 860 | Expression of Concern | 3 |
| English Abstract | 14173 | Autobiography | 825 | Overall | 3 |
| Clinical Trial, Phase II | 12344 | Technical Report | 815 | Legislation | 1 |

*Table 2: PT tag counts in selected corpus data.*

Additionally, multi-label classification introduces complexity, as MEDLINE citations often carry multiple PT tags [2], requiring models to predict both individual tags and their valid combinations. The predictive models must capture these multi-label dependencies effectively. Figure 1 shows the correlations of PT tags in our data.

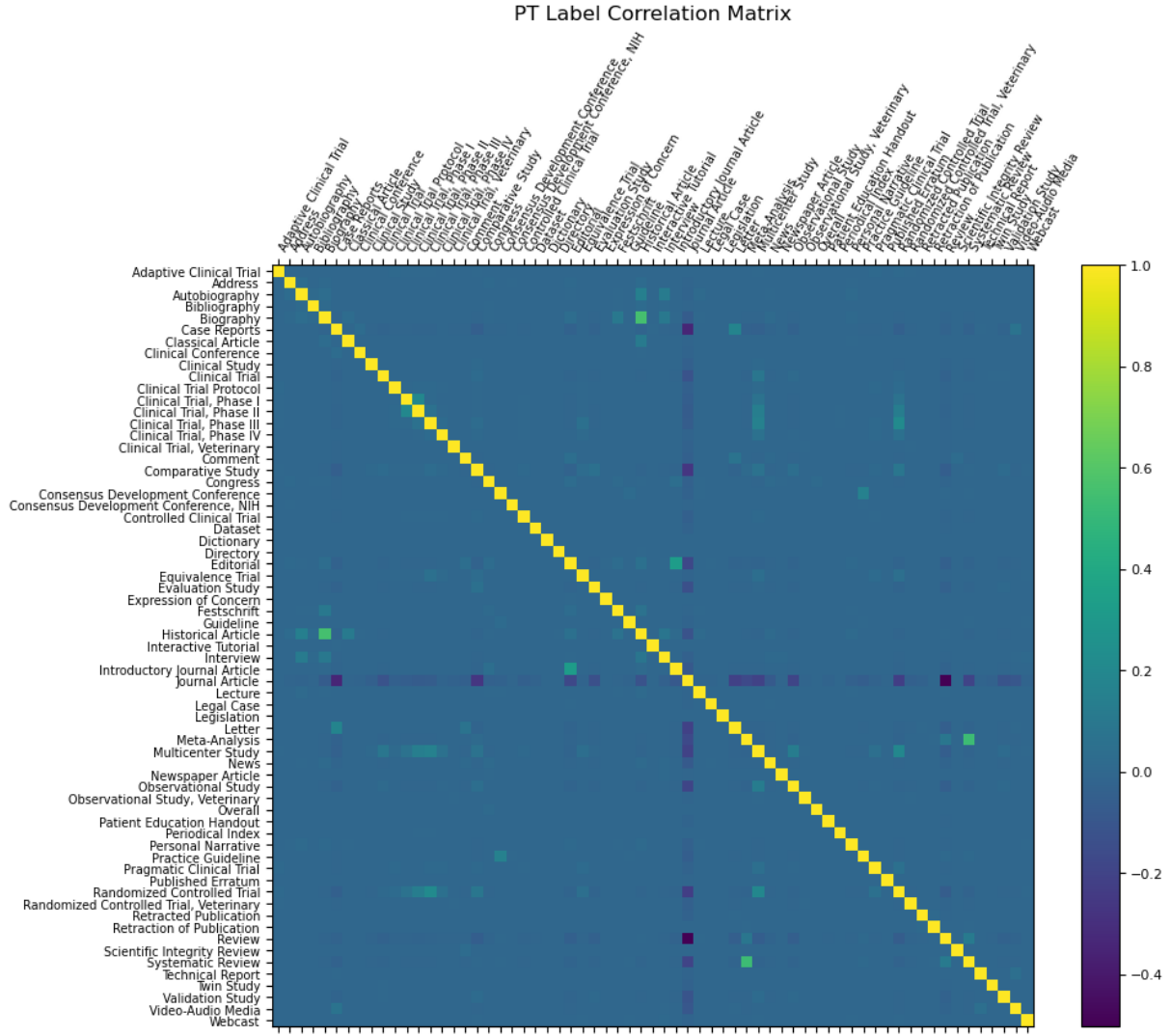

*Figure 1: MDELINE Publication Type tag correlations.*

Further complicating matters is the limited availability of consistently discriminative features. Key metadata, such as abstracts and titles, may be incomplete or insufficiently informative for nuanced tags, and unspecific PTs are particularly difficult to infer from metadata alone. Indexers traditionally examine full-text publications to determine PT tags, but our project requires us to rely solely on MEDLINE citation data.

Modeling Complexities: PT prediction is inherently complex due to the requirements of multi-label classification and adherence to NLM-developed co-occurrence and priority rules for PT tags [29]. While Transformer-based models like BERT and DistilBERT offer significant promise, they must manage the wide range of possible tag combinations and account for highly variable input lengths. Data preprocessing and feature engineering are critical to optimizing model performance. Truncating long inputs to meet the 512-token limit of the Transformer models considered in this study risks losing valuable contextual information, and balancing features like journal identifiers, titles, and abstracts requires careful prioritization to enhance prediction accuracy. Evaluation and optimization add another layer of complexity, as achieving a balance between high recall and acceptable precision is crucial to ensure comprehensive and accurate indexing.

Resource Constraints: The computational demands of training and fine-tuning large foundation models are substantial, requiring high-performance infrastructure to handle extensive datasets and complex multi-label tasks. Initial experiments often had to prioritize computationally efficient approaches due to restricted access to advanced hardware. Moreover, any proposed solution must be scalable to MEDLINE's daily influx of approximately up to 20,000 citations, requiring models that not only achieve high accuracy but also integrate seamlessly into existing workflows without introducing delays. The dynamic nature of PT tags, with periodic updates to the MeSH vocabulary, adds the additional burden of maintaining model adaptability and consistency over time.

## 4. Methodology

We used two distinct model architectures: monolithic multilabel classifiers and an ensemble of binary classifiers. Both required similar data preprocessing steps, with some differences as described below.

### 4.1. Data Sources and Preparation

4.1.1. Data Sources

The dataset for this research was derived from MEDLINE, an indexed subset of PubMed, which houses millions of biomedical citations. The study focused on citations indexed between January 2016 and March 2023 to ensure consistency in the application of MeSH Publication Types (PTs). This period was selected due to the relative stability in PT tagging conventions and the manual indexing practices employed during this time, which provide high-quality labeled examples. Citations indexed by automated systems were excluded to avoid potential inconsistencies introduced by legacy automation processes [30]. Prior to April 1, 2023, the NLM partially and fully automated the indexing of some journals, with humans reviewing selected citations. However, from that date onwards, the process became fully automated.

The corpus comprised over five million citations, each annotated with one or more PTs. However, only 66 distinct PT tags were included in the study, as unspecific PTs—such as "Preprint" or funding-related tags such as "Research Support, N.I.H., Intramural"—were excluded due to their dependence on external metadata or full-text information unavailable in the dataset. Additional tags were excluded due to other practical considerations, as explained below.

4.1.2. Data Cleaning

Data cleaning involved several critical steps to ensure that the dataset was both standardized and compatible with the machine learning models. Special formatting characters, HTML tags, and encoding anomalies were removed or replaced with their ASCII equivalents to streamline the text. For example, symbols such as "©" and "±" were converted to textual equivalents like "(c)" and "+/-". This preprocessing ensured that the data conformed to the expectations of the model tokenizers while preserving semantic meaning.

Punctuation was retained to aid the model's contextual understanding, but unnecessary or non-discriminative metadata fields were removed. Features such as copyright notices, date of publication, and publishing model indicators among others were excluded due to their limited relevance for PT prediction.

4.1.3. Feature Selection and Engineering

Feature selection was guided by the need to balance discriminative value with input length constraints imposed by Transformer models. Three citation metadata fields—Title, Abstract, and Journal ID—were identified as the most valuable features for prediction. Titles and Abstracts often provide direct or implicit clues about the PT, while Journal ID (formally called "NLM Unique ID") serves as a proxy for the journal title, offering insights into publication preferences while conserving model input length for other features. Other features explored include author names and affiliations, full journal titles, and publisher-provided keywords, but they subtracted from our token budget resulting in truncated abstracts and did not improve predictions. Also, only a fraction of citations have keywords.

To accommodate the 512-token model input limit, a structured input format was adopted that combined the selected citation features:

Journal ID<1>Title<2>Abstract

This format allowed models to infer the presence or absence of features, such as missing abstracts, from the input structure itself. Truncation prioritized retaining the beginning of the abstract, which typically contains the most relevant information, while ensuring that shorter fields, like the title and Journal ID, were preserved after truncation.

4.1.4. Data Partitioning and Balancing

Given the extreme class imbalance in the dataset, stratification and balancing were critical to ensure robust model performance [31] [32]. We allocated 90% of our data for training, 5% for evaluating the model during training, and 5% was held-off for testing the model after training. This partitioning was critical to ensure that our models could generalize well to unseen data. For binary classifiers, data balancing techniques included oversampling and subsampling classes and using class weights to penalize misclassifications. Each training, evaluation and test dataset for each model was designed to contain 50% positive examples and 50% negative examples. The negative examples were drawn from instances belonging to all other classes. This stratified sampling technique was essential to enhance the models' ability to recognize classes beyond the specific class each binary classifier was trained for.

For monolithic multi-label models, we ensured that the datasets included examples of all classes. The stratified sampling algorithms of Merilles and Du [33] was employed to maintain proportional representation of each PT tag across all partitions. This approach allowed us to effectively manage the widely varying number of examples available for each class within the corpus. Table 3 shows an example of our partitioning of the corpus for training and evaluating a multilabel model.

| Publication Type tag | Train | Evaluation | Test | Train% | Eval% | Test% |
|---|---|---|---|---|---|---|
| Journal Article | 3149302 | 156609 | 155659 | 90.98 | 4.52 | 4.50 |
| Review | 578756 | 28772 | 28519 | 90.99 | 4.52 | 4.48 |
| Case Reports | 286904 | 14251 | 14169 | 90.99 | 4.52 | 4.49 |
| Comparative Study | 174052 | 11098 | 11207 | 88.64 | 5.65 | 5.71 |
| Randomized Controlled Trial | 129762 | 6481 | 6524 | 90.89 | 4.54 | 4.57 |
| Letter | 112068 | 5688 | 5510 | 90.92 | 4.61 | 4.47 |
| Multicenter Study | 110179 | 5488 | 5535 | 90.91 | 4.53 | 4.57 |
| Systematic Review | 97789 | 4770 | 4850 | 91.04 | 4.44 | 4.52 |
| Observational Study | 89627 | 4425 | 4448 | 90.99 | 4.49 | 4.52 |
| Editorial | 79554 | 3846 | 3913 | 91.11 | 4.40 | 4.48 |
| Meta-Analysis | 75961 | 3713 | 3780 | 91.02 | 4.45 | 4.53 |
| Evaluation Study | 46586 | 2330 | 2391 | 90.80 | 4.54 | 4.66 |
| Clinical Trial | 35509 | 1776 | 1819 | 90.81 | 4.54 | 4.65 |
| Historical Article | 35177 | 1660 | 1723 | 91.23 | 4.30 | 4.47 |
| Validation Study | 31471 | 1544 | 1515 | 91.14 | 4.47 | 4.39 |
| Video-Audio Media | 22249 | 1102 | 1155 | 90.79 | 4.50 | 4.71 |
| Introductory Journal Article | 18475 | 914 | 890 | 91.10 | 4.51 | 4.39 |
| News | 16017 | 760 | 737 | 91.45 | 4.34 | 4.21 |
| Clinical Trial, Phase II | 11155 | 575 | 614 | 90.37 | 4.66 | 4.97 |
| Biography | 10899 | 513 | 555 | 91.08 | 4.29 | 4.64 |

*Table 3: PT tag count and percentages of top 25 tags after partitioning corpus into multilabel train, evaluation and test datasets using the Merrillees and Du's algorithm.*

This meticulous data balancing and partitioning strategy was instrumental in mitigating the effects of data imbalance, thereby enabling our models to achieve improved performance and generalization capabilities.

4.1.5. Corpus Normalization

Normalization efforts removed redundant "Journal Article" tags from citations already labeled with more specific PTs. This process ensured that the dataset emphasized meaningful distinctions between PTs, improving model learning and reducing bias toward dominant classes. Normalization extended beyond cleaning and balancing to address the diversity of PT combinations. Citations with multiple PT tags were analyzed to understand common co-occurrence patterns, enabling the development of heuristics for tag compilation and rule-based filtering. This step ensured that predictions aligned with MEDLINE's indexing standards, further enhancing the models' practical applicability.

**4.2. Predictive Application Architectures**

The study involved fine-tuning state-of-the-art, pre-trained Transformer-based machine learning models on labeled MEDLINE citation metadata and developing prototype applications to infer PT tags. Two primary architectures were explored: monolithic multi-label classifiers and ensembles of binary classifiers.

4.2.1. Monolithic Multi-Label Classifier

The multi-label classifier architecture used a single foundation model to predict multiple PT tags simultaneously for each citation (see Figure 2). This approach exploited the ability of modern Transformer architectures to understand the relationships between different PTs, allowing for predictions that captured co-occurrence patterns directly from the data. A 'Tag compiler" module allowed optimizing the final list of predicted PT tags.

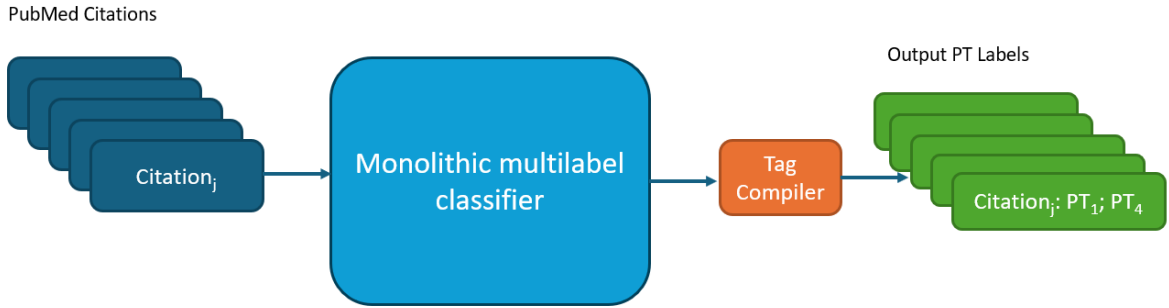

*Figure 2: Monolithic multilabel classifier architecture.*

Advantages:

- Efficiency: By handling all PT predictions in one pass.
- Integrated Context: The model leveraged contextual relationships between PTs, improving performance on citations with multiple tags.
- Simplified Maintenance: A single-model solution is easier to train (or fine-tune), deploy, and maintain compared to multi-model architectures.

Limitations:

- Data Demand: Achieving high accuracy required large, balanced datasets that represented all PT tags adequately.
- Less Granular Control: The unified approach made it challenging to isolate and improve predictions for specific underperforming PTs.

4.2.2. Ensemble of Binary Classifiers

This approach aimed to optimize the classification performance of individual PT tags by training separate binary classifiers, each dedicated to predicting the presence or absence of a specific PT tag for a given citation (see Figure 3). For example, one classifier focused solely on identifying "Systematic Review", while another handled "Randomized Controlled Trial". The ensemble architecture allowed for greater control over the prediction of individual PTs, making it easier to optimize for rare tags by applying unique training strategies for each classifier.

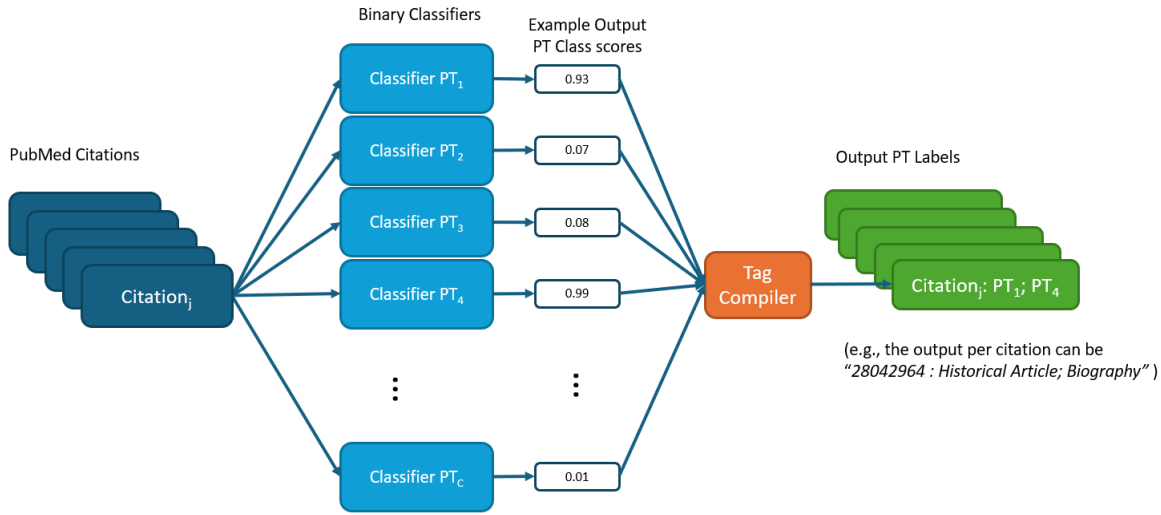

*Figure 3: Ensemble of binary classifiers architecture, with example PT class recall scores.*

Advantages:

- Customization: Classifiers could be fine-tuned independently to handle imbalanced class distributions effectively.
- Flexibility: Poorly performing classifiers for certain PTs could be re-trained or replaced without affecting others.
- Modular Tagging: This design facilitated compliance with MEDLINE's PT tag co-occurrence and priority rules by allowing finer control of predictions through a "tag compiler" module.

Limitations:

- Scalability: Processing a citation required running it through multiple classifiers sequentially, leading to longer inference times.
- Resource Intensity: Training and maintaining dozens of binary classifiers demanded significant computational resources, especially for rare PTs.

4.2.3. Pre-Trained Foundation Models

Both approaches relied on fine-tuning pre-trained foundation models to adapt them to the domain-specific task of PT prediction. The models explored include:

BERT (Bidirectional Encoder Representations from Transformers): A widely used model known for its bidirectional understanding of text, BERT provided a strong baseline for the task. The BERT-base that we utilized has proven to be capable of handling large, complex datasets. This model has 110 million parameters, 12 transformer layers, a hidden size of 768, and 12 self-attention heads.

DistilBERT: A lighter, faster version of BERT that retains much of BERT's performance while being more computationally efficient. This makes it an excellent choice for initial experiments and resource-constrained scenarios. It has 40% less parameters than BERT-base-uncased, has shown to run 60% faster while preserving over 97% of BERT's performance as measured on widely used benchmarks [43].

Other models: Several Transformer models were initially considered for this study, including versions of RoBERTa and BioBERT, but preliminary tests revealed no significant performance advantages of the alternatives over BERT in terms of key metrics, such as precision, recall, and F1-score, for the PT prediction task. Given these early observations and considering the resource-intensive nature of fine-tuning and evaluating multiple Transformer models across the large-scale MEDLINE dataset, we prioritized BERT and DistilBERT. By concentrating on these two models, we ensured a comprehensive exploration of model performance within practical constraints, while the insights gained remain broadly applicable to other Transformer-based architectures in similar tasks.

Cased vs. Uncased Versions: Cased versions of the models were used to preserve capitalization, which is critical in biomedical texts where case often distinguishes entities (e.g., "DNA" vs. "dna").

4.2.4. Model Selection and Fine-Tuning

The choice of model and architecture was guided by the specific demands of the task and the constraints of our computational resources. Fine-tuning the 110-million-parameter BERT model on a limited hardware budget posed significant challenges due to the complexity of our dataset: an extensive vocabulary (> 35,000 words), long sequence lengths (512 tokens), and over 5 million input citations. These factors placed considerable strain on the available resources, particularly for experiments involving an ensemble of binary classifiers. Fine-tuning a large number of BERT-based classifiers independently was prohibitively time- and resource-intensive, with each model requiring from hours to days to fine-tune. To address these limitations, we turned to DistilBERT for the ensemble solution, capitalizing on its efficiency and reduced computational requirements while still achieving competitive performance. This approach enabled us to scale the ensemble method effectively within our resource constraints, complementing the use of BERT in our monolithic multi-label experiments.

Fine-tuning involved adjusting the pre-trained models to recognize patterns in MEDLINE citation metadata. Training datasets were stratified and balanced to address class imbalance, and hyperparameters such as learning rate, batch size, and regularization weights were optimized for each model.

4.2.5. Tag Compilation Algorithm

A "tag compiler" module was employed in both solution architectures to refine model outputs. This module filtered low-confidence predictions, ensured adherence to tag co-occurrence rules, and limited the number of predicted tags based on corpus statistics or model-derived estimates.

### 4.3. Evaluation Metrics

To rigorously evaluate the performance of the machine learning models used for PT prediction, several well-established metrics were employed. These metrics were selected to address the multi-label and highly imbalanced nature of the task, ensuring a comprehensive assessment of model accuracy, precision, recall, and overall effectiveness in real-world indexing scenarios.

Two AUC metrics were employed during fine-tuning to evaluate the models' performance, especially in the context of imbalanced datasets: AUC-ROC (Receiver Operating Characteristics) to evaluate the trade-offs between true positive and false positive rates across decision thresholds, and AUC-PR (Precision-Recall) in scenarios with class imbalance [34] [35].

After fine-tuning, precision, recall, and F1-score were the primary metrics for evaluating the performance of individual PT predictions with our hold-off datasets, especially for binary classifiers. For multi-label classifiers, we used macro-averaged Precision, Recall and F1-Score, which treats each PT class equally, averaging metrics across all classes regardless of their frequency.

Given the probabilistic nature of model outputs, thresholds for tag inclusion in result sets were optimized by our "tag compiler" modules to balance precision and recall:

- Fixed Threshold: A universal threshold (e.g., 0.5) for all PT tags.
- Dynamic thresholding: Thresholds tuned per PT tag based on corpus statistics and model performance during validation.

In our assessment of the practical applicability of our predictive approaches in MEDLINE's indexing workflows, these aspects were of particular interest:

- Tag sensitivity: the application's ability to recognize relevant tags, even at the cost of some false positives (i.e., we prioritized recall over precision).
- Tag List Length Predictions: The application's ability to predict the appropriate number of PT tags for each citation, adhering to corpus statistics.
- Tag ruled adherence: compliance with MEDLINE's co-occurrence and priority rules.

### 4.4. Experimental Setup

The experimental setup for this study was significantly shaped by resource constraints, requiring careful planning and prioritization to maximize the utility of limited computational infrastructure. The project faced challenges in securing consistent access to high-performance computing (HPC) resources, requiring strategic decisions regarding model training, data processing, and evaluation. This experimental setup reflects the pragmatic balance between ambition and resource availability, leveraging creative strategies to address computational limitations while advancing the feasibility of PT prediction using state-of-the-art machine learning models.

### 4.4.1. Consistent Software Environment

Although this study was executed across different computing platforms, all of them employed the same software setup, consisting of the Linux Ubuntu operating system, Python, and Jupyter Notebooks for scripting, model training, and interactive analysis. The TensorFlow-based *ktrain* Python library was used for most training, testing, and evaluation tasks. While slightly different versions of the software were used across platforms, their functionality remained consistent, ensuring compatibility and uniformity throughout the research process.

### 4.4.2. Cloud Computing

Cloud computing was tested as a potential resource to expand capabilities for this project; however, it was quickly determined that the costs associated with training large models on commercial platforms like AWS or Google Cloud were prohibitive within the project's budget. Given the need to process millions of citations and explore various configurations, the expenses for sustained cloud use proved unsustainable.

### 4.4.3. Local Computing Resources

At the project's outset, the available infrastructure consisted of a workstation equipped with an Intel 8-Core i7 CPU, 32 GB of RAM, and two NVIDIA RTX GPUs with 12 GB VRAM each. While sufficient for preliminary experiments, this setup imposed significant limitations on the scope and scale of initial model training. The limited VRAM of the GPUs required small batch sizes and careful management of input data to avoid memory overflow. In the later stages of the project, additional resources became available, including a server with a faster 16-Core CPU, 64 GB of RAM, and two NVIDIA RT 1080 Ti GPUs with 11GB VRAM each, as well as access to a local NVIDIA DGX1 machine featuring 40 CPU cores, 512 GB of RAM, and 8 x 12 GB NVIDIA GPUs. While still constrained compared to commercial cloud environments, these resources provided a critical boost in processing capacity.

### 4.4.4. Access to Biowulf

Biowulf is the National Institutes of Health's high performance computing cluster [36]. Midway through the project, limited access to the Biowulf cluster allowed experimentation with resource-intensive tasks. As a high-performance computing resource with advanced GPUs, expanded CPU cores, and higher RAM, Biowulf was extremely useful for fine-tuning larger models and processing extensive datasets. However, its shared use among many researchers led to limited time allocations and scheduling delays, necessitating careful planning to prioritize high-impact experiments.

4.4.5. Efficient Resource Allocation

Given the limitations, the experimental setup was designed to optimize resource usage:

- Model Selection: Although BERT and other large models were of significant interest, DistilBERT was prioritized for most models due to its smaller size and faster training time. This allowed for rapid experimentation and iteration within the constraints of the available GPU resources
- Data Subsampling: For preliminary experiments, data subsets were created by sampling citations to reduce memory and compute requirements while maintaining representative distributions of PT tags.
- Incremental Training: Models were fine-tuned incrementally, with larger datasets and hyperparameter tuning reserved for phases when more resources where available.
- Staggered Experiments: Tasks were scheduled strategically to align with resource availability, ensuring that compute-heavy experiments were conducted on Biowulf or upgraded servers, while lighter tasks were conducted in our smaller computers.

4.4.6. Resource Constrain Mitigation

Even with upgraded computing resources, fine-tuning large models such as RoBERTa or BERT-large on the entire dataset proved infeasible within the project timeline. A key concern for future production models addressing the PT tag prediction problem is the need for periodic updates and fine-tuning of the predictive models, both to align with evolving indexing standards and to support ongoing model improvement efforts. These updates can leverage additional data to enhance performance for certain PT classes and enable the system to handle an expanded range of PT tags, maintaining its utility and effectiveness over time.

To mitigate resource constraints, techniques such as mixed precision quantization were explored as potential solutions to reduce computational demands and remain as an attractive option for future work. However, mixed precision support on the computing platforms available at the time was not fully stable, introducing risks of suboptimal results and adding complexity to the training process. These challenges, combined with the time and cost associated with addressing potential issues, made full precision a more practical and reliable choice for this pilot project. Furthermore, while mixed precision can enhance training performance, it may compromise model accuracy [37]. Given that maximizing accuracy was a critical objective of this study, all models were trained in full precision to ensure the best possible performance.

## 5. Results

The experiments provided valuable insights into the performance of selective predictive models and the effectiveness of various strategies for addressing challenges in PT prediction. Below, we elaborate on the findings for the major types of models used and the role of stratified sampling in enhancing performance.

### 5.1. Monolithic Multi-Label Classifier

The monolithic models demonstrated robust performance, particularly for the most frequent PT classes in the corpus, effectively capturing co-occurrence patterns and aligning with corpus statistics. To evaluate scalability, models fine-tuned to handle varying numbers of PT tags were tested, with Table 4 presenting an illustrative example.

| Publication Type | Precision | Recall | F1-score |
|---|---|---|---|
| Journal Article | 0.99 | 0.99 | 0.99 |
| Review | 0.88 | 0.87 | 0.88 |
| Case Reports | 0.91 | 0.90 | 0.91 |
| Comparative Study | 0.64 | 0.46 | 0.53 |
| Randomized Controlled Trial | 0.85 | 0.81 | 0.83 |
| Letter | 0.77 | 0.75 | 0.76 |
| Multicenter Study | 0.78 | 0.54 | 0.64 |
| Systematic Review | 0.93 | 0.90 | 0.91 |
| Observational Study | 0.77 | 0.52 | 0.63 |
| Editorial | 0.72 | 0.65 | 0.68 |

*Table 4: Performance results of top-10 PT tags monolithic BERT multilabel classifier.*

The performance of the multi-label PT tag classifiers exhibited a decline as the number of PT classes included in the fine-tuning process increased. This trend is attributed to the increased complexity of the classification task, where the model must not only identify individual PTs but also account for the relationships and co-occurrence patterns between a larger number of classes.

As the number of PT classes grows, the likelihood of incorrect predictions—either through false positives or false negatives—rises, leading to a broader distribution of errors across the expanded label set. Furthermore, the imbalance in class representation becomes more pronounced as rarer PTs are included, amplifying the challenges associated with learning effective representations for underrepresented classes. Table 5 illustrates this performance trend.

| Top-n Multi-label Model | Avg. Precision | Avg. Recall | Avg. Macro-Average F1-score |
|---|---|---|---|
| Top-10 PT tags | 0.83 | 0.74 | 0.78 |

| Top-15 PT tags | 0.67 | 0.72 | 0.69 |
| Top-52 PT classes | 0.36 | 0.51 | 0.34 |

Table 5: Average performance of multi-label BERT classifiers trained with top-n PT tags. The numbers shown are the micro-average for the corresponding metric across all the PT classes handled by each model.

This observation underscores the trade-offs inherent in multi-label classification tasks, particularly in domains like MEDLINE indexing, where the diversity of labels and their overlapping nature add layers of complexity. Future work could explore strategies to mitigate this performance degradation, such as dynamic weighting of classes, transfer learning with pre-trained models specialized in biomedical contexts, or hybrid architectures that balance generalizability with class-specific optimization.

**5.2. Ensemble of Binary Classifiers**

The ensemble approach, comprising individual binary classifiers for each PT, demonstrated strong performance for classes with abundant data, such as "Journal Article" and "Review." Precision, recall, and F1-scores for these dominant tags were consistently high, often exceeding 0.90. For instance:  The classifier for "Journal Article" achieved an F1-score of 0.89, with a recall of 0.90, indicating reliable identification of this tag despite its dominance in the dataset. Rare PTs such as "Bibliography" showed reduced performance results (recall 0.83 but a F1-score of only 0.45). Extremely rare tags were excluded due to insufficient training examples in the corpus.

| # | Publication Type | Precision | Recall | F1-score |
|---|---|---|---|---|
| 1 | Journal Article | 0.88 | 0.90 | 0.89 |
| 2 | Review | 0.95 | 0.93 | 0.94 |
| 3 | Case Reports | 0.97 | 0.97 | 0.97 |
| 4 | Comparative Study | 0.88 | 0.89 | 0.88 |
| 5 | Randomized Controlled Trial | 0.93 | 0.92 | 0.93 |
| 6 | Letter | 0.95 | 0.97 | 0.96 |
| 7 | Multicenter Study | 0.89 | 0.89 | 0.89 |
| 8 | Systematic Review | 0.98 | 0.97 | 0.98 |
| 9 | Observational Study | 0.91 | 0.94 | 0.93 |
| 10 | Editorial | 0.92 | 0.95 | 0.93 |
| 11 | Meta-Analysis | 0.98 | 0.97 | 0.97 |
| 12 | Evaluation Study | 0.89 | 0.94 | 0.91 |
| 13 | Clinical Trial | 0.86 | 0.90 | 0.88 |
| 14 | Historical Article | 0.92 | 0.94 | 0.93 |
| 15 | Validation Study | 0.95 | 0.96 | 0.96 |
| 16 | Video-Audio Media | 0.91 | 0.94 | 0.93 |
| 17 | Introductory Journal Article | 0.93 | 0.93 | 0.93 |
| 18 | News | 0.95 | 0.94 | 0.95 |
| 19 | Clinical Trial, Phase II | 0.95 | 0.95 | 0.95 |
| 20 | Biography | 0.91 | 0.95 | 0.93 |
| 21 | Clinical Trial, Phase III | 0.98 | 0.93 | 0.95 |
| 22 | Practice Guideline | 0.93 | 0.98 | 0.95 |
| 23 | Clinical Trial, Phase I | 0.95 | 0.97 | 0.96 |
| 24 | Clinical Trial Protocol | 0.99 | 0.99 | 0.99 |
| 25 | Comment | 0.93 | 0.95 | 0.94 |

| | | | | | |
|---|---|---|---|---|---|
| 26 | Controlled Clinical Trial | 0.84 | 0.92 | 0.88 |
| 27 | Clinical Study | 0.79 | 0.93 | 0.86 |
| 28 | Personal Narrative | 0.89 | 0.91 | 0.90 |
| 29 | Congress | 0.99 | 0.89 | 0.90 |
| 30 | Interview | 0.95 | 0.92 | 0.93 |
| 31 | Randomized Controlled Trial, Veterinary | 0.93 | 0.99 | 0.96 |
| 32 | Consensus Development Conference | 0.91 | 0.94 | 0.92 |
| 33 | Twin Study | 0.99 | 0.96 | 0.98 |
| 34 | Pragmatic Clinical Trial | 0.99 | 0.88 | 0.89 |
| 35 | Clinical Trial, Veterinary | 0.96 | 0.99 | 0.98 |
| 36 | Dataset | 0.99 | 0.99 | 0.99 |
| 37 | Clinical Trial, Phase IV | 0.79 | 0.93 | 0.86 |
| 38 | Patient Education Handout | 0.98 | 0.95 | 0.96 |
| 39 | Equivalence Trial | 0.98 | 0.88 | 0.93 |
| 40 | Guideline | 0.95 | 0.95 | 0.95 |
| 41 | Autobiography | 0.81 | 0.93 | 0.87 |
| 42 | Technical Report | 0.84 | 0.93 | 0.88 |
| 43 | Webcast | 0.95 | 0.95 | 0.95 |
| 44 | Clinical Conference | 0.88 | 0.76 | 0.82 |
| 45 | Lecture | 0.93 | 0.72 | 0.81 |
| 46 | Classical Article | 0.94 | 0.89 | 0.91 |
| 47 | Address | 0.81 | 0.71 | 0.76 |
| 48 | Legal Case | 0.94 | 0.71 | 0.81 |
| 49 | Observational Study, Veterinary | 0.87 | 0.93 | 0.90 |
| 50 | Festschrift | 0.88 | 0.78 | 0.82 |
| 51 | Published Erratum | 0.78 | 0.78 | 0.78 |
| 52 | Bibliography | 0.31 | 0.83 | 0.45 |

Table 6: Evaluation Performance of top-52 DistilBERT binary classifiers fine-tuned with stratified data, sorted by PT tag prevalence in corpus.

We tested the ensemble's multi-label performance by limiting the maximum number of tags per citation to between 1 and 6 PT tags. We also filtered the predicted PT tags by recall values varying between 0.5 and 0.9, in 0.1 steps. We tested ensembles with binary classifiers for the top-n most frequent PT classes in the corpus, with n varying between 10 and 50. Truncation was performed at the end of lists of predicted tags, which were sorted by PT tag prevalence in the corpus, therefore this method removed excessive tags that were less likely to appear in the citations. Notably, in many cases, the additional filtering steps have already reduced the number of inferred PT tags to below the desired limit and for instance truncating the list of predicted tags was not always necessary. The top 4 performing ensembles are shown on Table 7.

| MAX TAGS | P-THRESHOLD | Num PT Classes | Cumulative Accuracy | Cumulative Precision | Cumulative Recall | Micro-Average F1 Score |
|---|---|---|---|---|---|---|
| 3 | 0.9 | 25 | 0.975 | 0.755 | 0.801 | 0.777 |
| 4 | 0.9 | 21 | 0.972 | 0.759 | 0.798 | 0.778 |
| 3 | 0.9 | 12 | 0.966 | 0.825 | 0.834 | 0.830 |
| 3 | 0.9 | 11 | 0.963 | 0.820 | 0.828 | 0.824 |

Table 7: Best performing ensembles of binary DistilBERT classifiers using fixed maximum numbers of PT tags per citation.

The primary advantage of this ensemble approach was its modularity, allowing each classifier to be independently optimized. However, the sequential processing required for inferencing across all classifiers introduced significant time overheads. Stratified sampling was crucial for this approach, ensuring that training datasets for each classifier included balanced proportions of positive and negative examples, which mitigated the effects of extreme class imbalance.

### 5.3. Inference Efficiency Comparison

The monolithic model's single-pass architecture is about 70% faster than the ensemble option, making it a more practical choice for real-world indexing workflows. Table 8 illustrates the differences in inferencing time of prototype applications making use of ensemble and monolithic models.

| Prototype Type | Avg. Time to Infer 20K Citations |
|---|---|
| Top-11 Ensemble - DistilBERT | 3.5 hours |
| Top-15 Ensemble - DistilBERT | 3.5 hours |
| Top-10 Monolithic – BERT base | 1 hour |
| Top-15 Monolithic – BERT-base | 1 hour |

*Table 8: Average inferencing time, rounded to closest half hour, of 5 ensembles and 10 monolithic runs, each predicting PT tags for 20,000 MEDLINE bibliographical references in a workstation with eight i7 cores and 32GB Ram.*

### 5.4. Impact of Stratified Sampling

Stratified sampling emerged as a cornerstone of the experimental strategy, addressing critical challenges posed by class imbalance and multi-label classification:

Improved recall for minority classes: In multi-label experiments, stratified datasets ensured that all combinations of PTs, including rare ones, were adequately represented, enhancing the model's ability to generalize. For example, for "Meta-Analysis," a minority/low frequency tag, the use of stratified sampling improved recall by almost 20% compared to non-stratified datasets, as the training process included proportionate examples of this tag. By ensuring adequate representation of rare PTs in training datasets, stratified sampling enhanced the model's ability to predict these tags accurately.

Balanced evaluation: Stratified test sets provided a fair and comprehensive assessment of model performance across all PTs, avoiding skewed results that overemphasize dominant tags.

Preservation of co-occurrence patterns: For the monolithic model, stratified sampling preserved the diversity of PT combinations, which was crucial for recognizing realistic patterns in multi-tagged citations.

## 6. Discussion

### 6.1. Interpretation of Results

The experimental results demonstrate that modern Transformer-based models can significantly enhance the accuracy and efficiency of PT tagging for MEDLINE citations. Both the ensemble of binary classifiers and the monolithic multi-label classifier showed notable improvements in recall compared to legacy systems, ensuring more comprehensive retrieval of biomedical literature. The ensemble approach excelled in precision, particularly for dominant PTs such as "Journal Article" and "Review," but its sequential processing demands made it less practical for high-throughput scenarios. In contrast, the monolithic model balanced precision and recall effectively, offering a more scalable solution capable of handling the daily influx of citations in MEDLINE workflows. However, this approach's performance weakened for less frequent PT tags and as more PT classes were added to the model.

The results also highlighted the importance of addressing class imbalance. Without strategies such as stratified sampling and weighted loss functions, low frequency PTs like "Adaptive Clinical Trial" and "Pragmatic Clinical Trial" would tend to be under-predicted. This finding underscores the need for careful data preparation and model optimization to ensure equitable performance across all PT classes.

### 6.2. Comparative Insights

The comparative analysis between the ensemble and monolithic approaches revealed distinct trade-offs that must be considered when selecting a method for deployment:

- Accuracy vs. Scalability: While the ensemble of binary classifiers achieved slightly higher precision for dominant PTs, its inference times scaled linearly with the number of PTs, making it less apt for real-time indexing. The monolithic model's single-pass architecture significantly reduced processing time while maintaining competitive accuracy.
- Granularity vs. Integration: The ensemble's modularity allowed for targeted optimizations of individual PTs, making it a good choice for scenarios where specific tags require higher accuracy. However, this modularity came at the cost of increased complexity in integrating the models into a unified workflow. Despite most of our binary classifiers exhibiting excellent performance individually, integrating them into our ensemble to predict multiple labels proved challenging. The monolithic model, by contrast, offered streamlined integration at the expense of fine-grained control over individual PT predictions.
- PT tagging Performance: The monolithic BERT model achieved strong performance when restricted to a limited number of PT tags but showed the challenges in managing the added complexity of predicting a broader range of PT tags. In contrast, the ensemble of binary classifiers exhibited consistent and competitive performance across configurations with varying numbers of PT classes and maximum tags per citation. It demonstrated its ability to handle greater diversity while preserving robust precision and recall.

These results indicate that while the monolithic model offers simplicity and efficiency for smaller tag sets, the ensemble approach provides superior flexibility and scalability for more complex PT prediction tasks, particularly in scenarios with diverse and multi-tagged citations.

These insights emphasize the need for a hybrid approach in some contexts, leveraging the strengths of both architectures to achieve optimal results.

**6.3. Lessons Learned**

Several key lessons emerged from this study, providing valuable guidance for future research and development in PT prediction:

The Critical Role of Data Preparation: Effective data preparation, including stratified sampling, normalization, and feature selection, proved essential for model performance. Ensuring balanced representation of PTs in training datasets addressed class imbalance, while carefully engineered input formats optimized the use of available metadata.

The Value of Scalable Models: The ability to process 20,000 citations daily is a non-negotiable requirement for MEDLINE workflows. While precision is important, scalability and inference speed are equally critical, highlighting the practicality of lightweight models such as DistilBERT and the efficiency of monolithic architectures.

Trade-Offs in Evaluation Metrics: The prioritization of recall over precision was a deliberate choice to enhance the utility of PT tags for information retrieval. This decision aligns with MEDLINE's goal of maximizing literature discovery but requires acceptance of higher false positive rates, which must be mitigated through post-processing algorithms like the tag compiler.

Adaptability of Transformer Models: The adaptability of pre-trained Transformer models to domain-specific tasks underscores their potential for broader applications in biomedical text classification. However, this adaptability comes with significant computational demands, necessitating access to advanced resources for fine-tuning and deployment.

Importance of Model Maintenance: The dynamic nature of MeSH vocabulary and PT definitions requires models to be periodically retrained and updated. This emphasizes the need for modular, maintainable solutions that can evolve alongside indexing standards.

Access to model training platforms: One of the key challenges encountered during this project was gaining access to appropriate hardware for training/fine-tuning complex machine learning models. While advances in hardware development—such as GPUs optimized for deep learning, cloud computing platforms, and high-performance computing clusters—have significantly improved the accessibility of training infrastructure, barriers remain in specific contexts. Additionally, competition for shared institutional resources, such as HPC clusters, further restricts access, leading to delays and constrained experimental timelines. These limitations highlight the need for strategic planning, resource allocation, and potentially the development of lightweight training strategies or hybrid models that balance performance with computational feasibility. Advances in affordable, energy-efficient hardware tailored for machine learning tasks could play a critical role in alleviating these constraints.

The findings of this study validate the potential of modern ML models to transform PT prediction, addressing longstanding challenges in scalability, accuracy, and resource efficiency. By balancing the trade-offs between precision, recall, and computational demands, this research offers actionable insights for advancing MEDLINE's automated indexing workflows. The lessons learned can help the application of practical solutions and open up new possibilities in bibliometric classification, leading to improvements in accessing biomedical knowledge.

## 7. Conclusion

This study demonstrates the feasibility and advantages of using Transformer-based machine learning models to predict MeSH Publication Types (PTs) for MEDLINE citations. The results highlight the substantial opportunities that these technologies offer to improve the automated indexing of MEDLINE citations.

### 7.1. Summary of Findings

The ensemble of binary classifiers and the monolithic multi-label classifier both showed strong potential for enhancing PT tagging:

Ensemble Approach: The modularity of binary classifiers allowed precise optimization for individual PTs, achieving F1-scores exceeding 0.90 for dominant tags like "Journal Article" and "Review." However, the ensemble's sequential inferencing demands posed challenges for scalability, particularly in high-throughput workflows.

Monolithic Multi-Label Model: By predicting multiple PTs simultaneously, the monolithic model demonstrated better efficiency while maintaining competitive performance across both dominant and rare PTs. Its macro and micro F1-scores of 0.75–0.85 indicated balanced capabilities, especially in recognizing co-occurrence patterns in multi-tagged citations.

Stratified sampling played a pivotal role in both approaches, addressing extreme class imbalance and ensuring robust performance across all PT classes. The tag compiler further refined model outputs by enforcing co-occurrence rules and optimizing tag list lengths, ensuring adherence to MEDLINE indexing standards.

### 7.2. Implications for Future Research

These findings offer several avenues for advancing PT prediction and automated biomedical indexing:

Incorporating Full-Text Data: Expanding the input features to include full-text articles could enhance model accuracy for nuanced PTs. This would address limitations associated with incomplete or ambiguous metadata.

Hybrid Architectures: A hybrid approach combining the ensemble's modular precision with the monolithic model's scalability could offer an optimal balance for real-world applications. For example, an ensemble could be deployed selectively for rare or high-priority PTs, while the monolithic model handles common tags.

Dynamic Adaptation to Vocabulary Changes: The evolving nature of MeSH PTs necessitates ongoing model updates. Future research should explore continual learning strategies to streamline model retraining and maintain alignment with indexing standards.

Efficiency Improvements: Further optimization of lightweight models like DistilBERT could make large-scale deployments more feasible, particularly for institutions with limited computational resources. Research into pruning, quantization, and other model compression techniques could enhance performance without sacrificing accuracy.

Broader Applications: The methodologies and findings from this study could extend to other bibliometric classification tasks, such as thematic tagging, citation analysis, and clustering, providing a foundation for further advancements in scientific literature indexing.

In conclusion, this study lays the groundwork for modernizing PT prediction in MEDLINE workflows, demonstrating the transformative potential of Transformer-based models. By addressing key challenges in accuracy, scalability, and adaptability, the research provides a promising path forward for enhancing access to biomedical knowledge and fostering innovation in bibliometric classification.

## 8. Acknowledgments

This work was supported by the Lister Hill National Center for Biomedical Communications of the National Library of Medicine (NLM), National Institutes of Health.